\documentclass[aps,prb,reprint,preprintnumbers,superscriptaddress,amsmath,amssymb,bibnotes,longbibliography]{revtex4-2}

\usepackage{graphicx}
\usepackage{dcolumn}
\usepackage{bm}
\usepackage[colorlinks,linkcolor=blue,anchorcolor=blue,citecolor=blue,urlcolor=blue,filecolor=blue,menucolor=blue,runcolor=blue]{hyperref}
\usepackage{amsmath} 
\usepackage{amssymb}
\usepackage{color}

\begin{document}
	
	
\title{Nodeless superconductivity in the topological nodal-line semimetal CaSb$_2$}
	
\author{Weiyin Duan}
\affiliation  {Center for Correlated Matter and Department of Physics, Zhejiang University, Hangzhou 310058, China}
\affiliation  {Zhejiang Province Key Laboratory of Quantum Technology and Device, Department of Physics, Zhejiang University, Hangzhou 310058, China}

\author{Jiawen Zhang}
\affiliation  {Center for Correlated Matter and Department of Physics, Zhejiang University, Hangzhou 310058, China}
\affiliation  {Zhejiang Province Key Laboratory of Quantum Technology and Device, Department of Physics, Zhejiang University, Hangzhou 310058, China}

\author{Rohit Kumar}
\affiliation  {Center for Correlated Matter and Department of Physics, Zhejiang University, Hangzhou 310058, China}
\affiliation  {Zhejiang Province Key Laboratory of Quantum Technology and Device, Department of Physics, Zhejiang University, Hangzhou 310058, China}

\author{Hang Su}
\affiliation  {Center for Correlated Matter and Department of Physics, Zhejiang University, Hangzhou 310058, China}
\affiliation  {Zhejiang Province Key Laboratory of Quantum Technology and Device, Department of Physics, Zhejiang University, Hangzhou 310058, China}

\author{Zhiyong Nie}
\affiliation  {Center for Correlated Matter and Department of Physics, Zhejiang University, Hangzhou 310058, China}
\affiliation  {Zhejiang Province Key Laboratory of Quantum Technology and Device, Department of Physics, Zhejiang University, Hangzhou 310058, China}

\author{Ye Chen}
\affiliation  {Center for Correlated Matter and Department of Physics, Zhejiang University, Hangzhou 310058, China}
\affiliation  {Zhejiang Province Key Laboratory of Quantum Technology and Device, Department of Physics, Zhejiang University, Hangzhou 310058, China}

\author{Chao Cao}
\affiliation  {Center for Correlated Matter and Department of Physics, Zhejiang University, Hangzhou 310058, China}
\affiliation  {Zhejiang Province Key Laboratory of Quantum Technology and Device, Department of Physics, Zhejiang University, Hangzhou 310058, China}

\author{Yu Song}
\email[Corresponding author: ]{yusong$_$phys@zju.edu.cn}
\affiliation{Center for Correlated Matter and Department of Physics, Zhejiang University, Hangzhou 310058, China}
\affiliation  {Zhejiang Province Key Laboratory of Quantum Technology and Device, Department of Physics, Zhejiang University, Hangzhou 310058, China}

\author{Huiqiu Yuan}
\email[Corresponding author: ]{hqyuan@zju.edu.cn}
\affiliation  {Center for Correlated Matter and Department of Physics, Zhejiang University, Hangzhou 310058, China}
\affiliation  {Zhejiang Province Key Laboratory of Quantum Technology and Device, Department of Physics, Zhejiang University, Hangzhou 310058, China}
\affiliation  {State Key Laboratory of Silicon Materials, Zhejiang University, Hangzhou 310058, China}
\affiliation  {Collaborative Innovation Center of Advanced Microstructures, Nanjing 210093, China}

\date{\today}

\date{\today}

\begin{abstract}
CaSb$_2$ is a topological nodal-line semimetal that becomes superconducting below 1.6~K, providing an ideal platform to investigate the interplay between topologically nontrivial electronic bands and superconductivity.
In this work, we investigated the superconducting order parameter of CaSb$_2$ by measuring its magnetic penetration depth change $\Delta \lambda (T)$ down to 0.07~K, using a tunneling diode oscillator (TDO) based technique. Well inside the superconducting state, $\Delta \lambda (T)$ shows an exponential activated behavior, and provides direct evidence for a nodeless superconducting gap. By analyzing the temperature dependence of the superfluid density and the electronic specific heat, we find both can be consistently described by a two-gap $s$-wave model, in line with the presence of multiple Fermi surfaces associated with distinct Sb sites in this compound. These results demonstrate fully-gapped superconductivity in CaSb$_2$ and constrain the allowed pairing symmetry.

\end{abstract}

\maketitle

\section{INTRODUCTION}
Topological superconductors (TSCs) have drawn major research efforts in recent years, due to both fundamental interest in the unconventional superconducting state and the potential for applications in topological quantum computation \cite{1RevModPhys.83.1057,2Sato_2017,3RevModPhys.80.1083,4Kitaev_2001}.
One route towards discovering TSCs is by turning materials with topologically nontrivial band structures into superconductors via various tuning parameters. For instance, superconductivity in Cu$_x$Bi$_2$Se$_3$ is realized by intercalating Cu into the topological insulator Bi$_2$Se$_3$ \cite{5PhysRevLett.104.057001,6RevModPhys.82.3045,7,8PhysRevX.8.041024},  whereas in the type-\uppercase\expandafter{\romannumeral2} Weyl semimetal MoTe$_2$ and the three-dimensional Dirac semimetal Cd$_3$As$_2$, superconductivity can be induced through the application of hydrostatic pressure \cite{9,10,11,12PhysRevLett.115.187001}.
Alternatively, intrinsic TSCs may be realized in topological semimetals (TSMs) that exhibit superconducting ground states, without additional tuning. In addition, such superconducting TSMs offer a fertile ground to probe the interplay between various topologically protected states and superconductivity \cite{63C5TC02373D}.

Nodal-line semimetals, in which band crossings form closed loops (nodal lines) rather than discrete points (as in Dirac or Weyl semimetals) in momentum space \cite{43PhysRevB.84.235126,44,45}, have received significant attention recently. Due to their unique nontrivial electronic topology, nodal-line semimetals can exhibit long-range Coulomb interactions, a large surface polarization charge \cite{46PhysRevB.93.035138,47,48PhysRevB.95.075138}, and torus-shaped Fermi surfaces (FSs) when the nodal line is close to the Fermi energy $E_{\rm F}$. If odd-parity superconductivity in a nodal-line semimetal is dominantly induced on such torus-shaped FSs, topological superconductivity can be realized \cite{49PhysRevB.97.094508}. Therefore, nodal-line semimetals with superconducting ground states are promising TSC candidates. However, since the nodal-line can easily become gapped due to spin-orbit coupling (SOC), an additional symmetry is required to protect the nodal-line against SOC \cite{50PhysRevB.92.081201,51PhysRevB.94.195109}. As a result of these constraints, reports on superconducting nodal-line semimetals have been limited thus far \cite{59,61,62}.



Recently, CaSb$_2$ is found to be a Dirac nodal-line semimetal with a superconducting ground state, and has been the focus of much interest \cite{18,56PhysRevB.105.184504,30PhysRevMaterials.4.041801}. A non-saturating giant magnetoresistance, and a metal-to-insulator-like transition followed by a saturating behavior are observed in transport measurements on CaSb$_2$ under applied magnetic fields \cite{18,56PhysRevB.105.184504}. These observations are similar to those found in other TSMs \cite{21PhysRevX.5.031023,22PhysRevB.94.121115,23,24,25,26,27,28,29PhysRevB.94.174411}, and can be ascribed to Dirac electrons from the Dirac nodal lines in CaSb$_2$. In addition, the Dirac nodal lines in CaSb$_2$ are protected by the nonsymmorphic symmetry, and are thus robust against SOC. 
These unique properties make CaSb$_2$ an excellent candidate to look for topological superconductivity and investigate the interplay between nontrivial electronic band topology and superconductivity. 

To make progress in these directions, an important property that needs to be experimentally clarified is the nature of the superconducting order parameter in CaSb$_2$. Theoretically, symmetry enforced nodal lines were suggested to be present in CaSb$_2$ if the gap symmetry belongs to the $B_g$ representation \cite{39PhysRevResearch.3.023086}. Experimentally, specific heat measurements down to 0.22~K in CaSb$_2$ single crystal uncover clear deviations from expectations for a simple BCS superconductor \cite{56PhysRevB.105.184504}, and the superconducting transition temperature $T_{\rm c}$ of pressurized CaSb$_2$ forms a peak around 3.1~GPa \cite{32PhysRevB.104.L060504}, which can be attributed to unconventional pairing. While these results suggest CaSb$_2$ may harbor an unconventional superconducting state, $^{121/123}$Sb-nuclear quadrupole resonance (NQR) measurements revealed a coherence peak below $T_{\rm c}$ in polycrystalline samples of CaSb$_2$, which instead suggests conventional $s$-wave superconductivity \cite{31}. 
To resolve these differences and elucidate the superconducting order parameter of CaSb$_2$, experiments highly sensitive to low-energy electronic excitations inside the superconducting state are needed.

In this work, the magnetic penetration depth and electronic specific heat of CaSb$_2$ single crystals were measured to probe its superconducting gap structure. The temperature dependence of the magnetic penetration depth change $\Delta\lambda(T)$ was measured down to 0.07 K, using a technique based on the tunneling diode oscillator (TDO), which is extremely sensitive to low-energy electronic excitations. The $\Delta \lambda (T)$ of CaSb$_2$ exhibits an exponential behavior and flattens at low temperatures, evidencing a nodeless superconducting ground state. By analyzing the electronic specific heat and the superfluid density derived from $\Delta\lambda(T)$, we find both results are consistently captured by a two-gap $s$-wave model. Our findings provide evidence for nodeless multi-gap superconductivity in the nodal-line semimetal CaSb$_2$.


\begin{figure}
	\centering
	\includegraphics[width=0.99\linewidth]{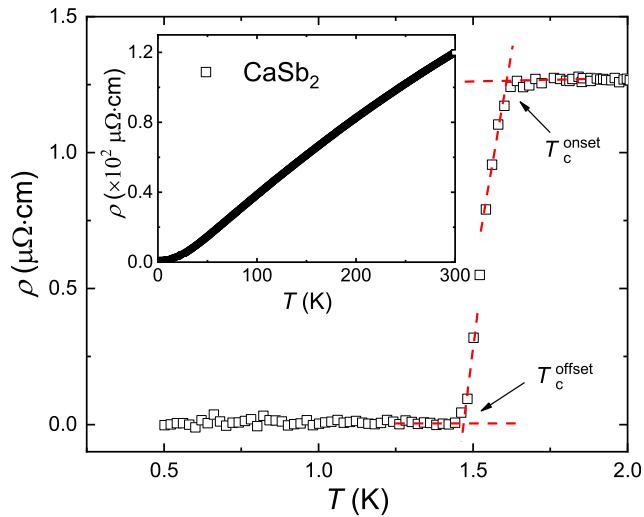} 
	\caption{The temperature dependence of the low temperature electrical resistivity $\rho (T)$ for CaSb$_2$, with a sharp superconducting transition at 1.6~K. The inset shows $\rho(T)$ from 0.5~K up to 300~K.}
	\label{Figure1}
\end{figure}

\section{Experimental Details}
Single crystals of CaSb$_2$ were grown using the Sb self-flux method, as described in Ref.~\cite{56PhysRevB.105.184504}. Electrical resistivity $\rho(T)$ was measured in a $^3$He cryostat using the four probe method. The specific heat was measured in a Quantum Design Physical Property Measurement System (PPMS) with a $^3$He insert. The temperature dependence of the magnetic penetration depth shift $\Delta\lambda(T) = \lambda(T) - \lambda(0)$ was measured using a TDO operating at about 7~MHz \cite{33doi:10.1063/1.1134272,34TDO2}. $\Delta\lambda (T)$ can be obtained from the TDO frequency shift  $\Delta f (T)$ through $\Delta \lambda (T) = G \Delta f(T)$, where $G$ is the calibration factor determined
from the geometry of the coil and the sample \cite{35PhysRevB.62.115}. TDO measurements down to 0.35~K and 0.07~K were respectively carried out in $^3$He and dilution refrigerators.
Two samples were separately mounted on a sapphire rod and inserted into the TDO coil without contact.
The rod is connected to a copper stage, where the thermometer is mounted. The coil of the TDO generates a very tiny ac magnetic field ($\approx 2$~$\mu$T) along the $c$-axis, which is much smaller than the lower critical field of CaSb$_2$ ($H_{\rm c1} \approx 3.9$ mT~\cite{56PhysRevB.105.184504}), ensuring the samples are in the full Meissner state once cooled slightly below $T_{\rm c}$. 

\begin{figure}
	\centering
	\includegraphics[width=0.99\linewidth]{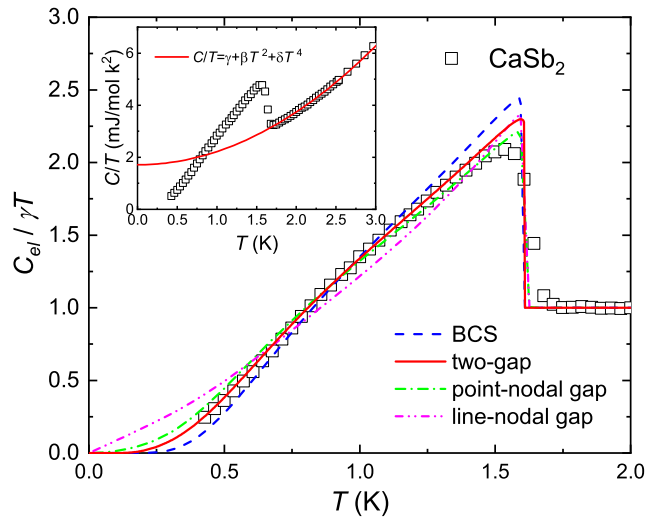} 
	\caption{The electronic specific heat $C_{\rm el} (T)/\gamma T$ for CaSb$_2$, with a clear superconducting transition at $T_{\rm c }=1.6$~K. The solid, dash, dash-dotted, and dash-dot-dotted lines are respectively fits to a two-gap $s$-wave model, a simple BCS model, a $p$-wave model with point nodes, and a $d$-wave model with line nodes. The inset shows the total specific heat $C(T)/T$, fit to $C(T)/T=\gamma+\beta T^2+\delta T^4$ in the normal state.}
	\label{Figure2}
\end{figure}

\section{Experimental Results and Discussion}
The inset of Fig.~\ref{Figure1} shows the electrical resistivity $\rho(T)$ of CaSb$_2$ from 300~K down to 0.5~K, revealing a metallic normal state with a residual resistivity of about 1.2~$\mu\Omega$~cm and a large residual resistivity ratio (RRR) of $\rho({\rm 300~K})/\rho({\rm 1.6~K}) = 95$, comparable with previous works \cite{56PhysRevB.105.184504,31}. Fig.~\ref{Figure1} zooms into $\rho(T)$ below 2~K, revealing a clear superconducting transition that onsets around 1.6~K and offsets around 1.46~K. Both the large RRR and the small transition width (0.14~K) demonstrate the samples used in this work are of high quality.

The total specific heat $C(T)/T$ of CaSb$_2$ is shown in the inset of Fig.~\ref{Figure2}, with a clear superconducting jump around $T_{\rm c}=1.6$~K which indicates the appearance of bulk superconductivity, consistent with electrical resistivity measurements. In the normal state, the specific heat can be modeled using $C(T)/T=\gamma+\beta~T^2+\delta~T^4$ (solid red line in the inset of Fig.~\ref{Figure2}), with $\gamma=1.71$~mJ~mol$^{-1}$K$^{-2}$, $\beta=0.507$~mJ~mol$^{-1}$K$^{-4}$ and $\delta=0.0079$~$\mu$J~mol$^{-1}$K$^{-6}$. $\gamma$ is the Sommerfeld coefficient, and the other two parameters characterize the phonon contribution to the specific heat. After subtracting the phonon contribution, the electronic specific heat $C_{\rm el}(T)/\gamma T$ is obtained, as shown in the main panel of Fig.~\ref{Figure2}. The simple BCS model, $p$-wave model with point nodes, and $d$-wave model with line nodes cannot capture the temperature evolution of $C_{\rm el}(T)/\gamma~T$, consistent with Ref.~\cite{56PhysRevB.105.184504}. On the other hand, a two-gap $s$-wave model gives an excellent description of the experimental data over the full temperature range (solid red line). The fit gap values are $\Delta_1(0)=1.77 k_{\rm B}T_{\rm c}$ and $\Delta_2(0)=0.88~k_{\rm B}T_{\rm c}$, with respective weights of 85\% and 15\%.

\begin{figure}
	\centering
	\includegraphics[width=0.99\linewidth]{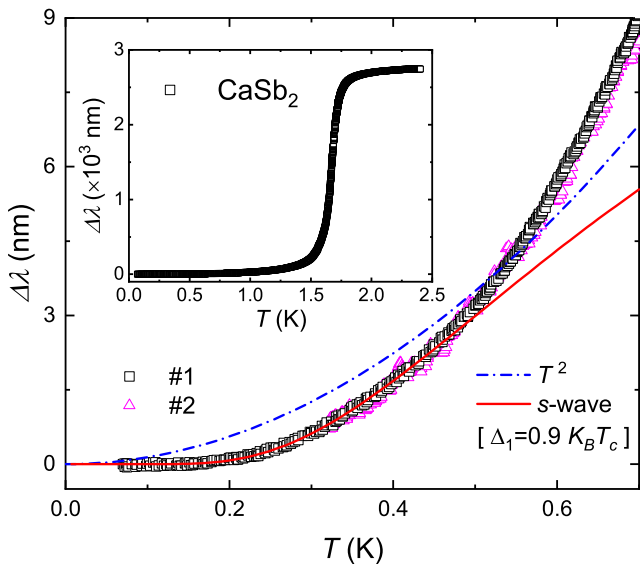} 
	\caption{The change of the magnetic penetration depth $\Delta\lambda(T)$, measured for two samples (\#1 and \#2). The dash-dotted and solid lines are respectively fits to the $T^2$ model and Eq.~\ref{equation1} at low temperatures. The inset shows $\Delta\lambda(T)$ from 2.5~K down to 0.07~K, with a clear superconducting transition observed around 1.6~K, consistent with electrical resistivity and heat capacity measurements.}
	\label{Figure3}
\end{figure}

\begin{figure}
	\centering
	\includegraphics[width=0.99\linewidth]{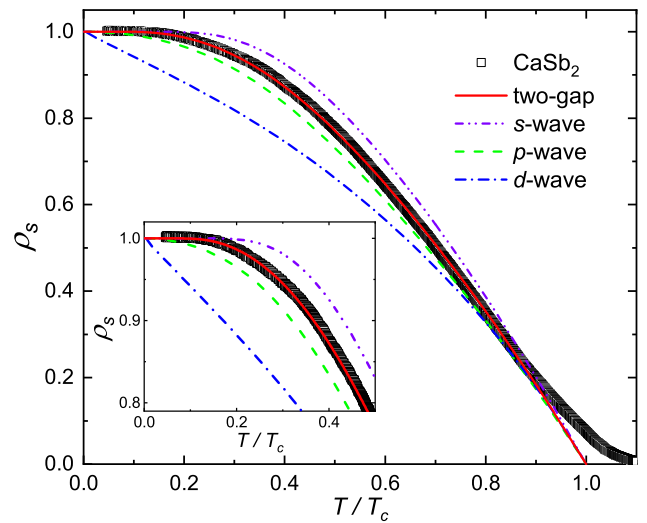} 
	\caption{Normalized superfluid density $\rho_{\rm s}(T)$ for sample \#1 as a function of the reduced temperature $T/T_{\rm c}$, obtained using $\lambda(0)~=~89.5$~nm. The solid, dash-dot-dotted, dashed, and dash-dotted lines are respectively fits to models with two $s$-wave gaps, a single $s$-wave gap, a $p$-wave gap, and a $d$-wave gap. The inset zooms into the low temperature region of $\rho_{\rm s}(T)$. 
	}
	\label{Figure4}
\end{figure}

To further probe the superconducting gap structure of CaSb$_2$, we measured the magnetic penetration depth change $\Delta\lambda(T)$ of two different samples, which are converted from the frequency shift $\Delta f(T)$ with respective calibration constants $G~=~11.0~\textrm{{\AA}/Hz}$ and $12.2~\textrm{{\AA}/Hz}$.
In Fig.~\ref{Figure3}, $\Delta\lambda(T)$ for CaSb$_2$ from 2.5~K down to 0.07~K is shown in the inset, with a clear reduction upon cooling around $T_{\rm c}=1.6$~K due to superconductivity, consistent with electrical resistivity and specific heat measurements in Figs.~\ref{Figure1} and \ref{Figure2}.
In the main panel of Fig.~\ref{Figure3}, $\Delta\lambda(T)$ is zoomed in to focus on its  behavior well below $T_{\rm c}$. Measurements for the two CaSb$_2$ samples show excellent agreement between each other.

Following the method in Ref.~\cite{57PhysRevB.19.4545}, using the Sommerfeld coefficient of $\gamma$ = 1.71 mJ mol$^{-1}$ K$^{-2}$, the residual resistivity of 1.2~$\mu\Omega$~cm obtained above, and a coherence length of $\xi_0=116$~nm from Ref.~\cite{56PhysRevB.105.184504}, we estimate the mean free path in CaSb$_2$ to be 947~nm. This is much larger than the coherence length, indicating that the sample is in the clean limit. 

In the clean limit, the magnetic penetration depth at low temperatures is expected to show $T$-linear or $T^2$ behavior in superconductors with line nodes or point nodes, respectively.  
As shown in Fig.~\ref{Figure3}, the $\Delta\lambda(T)$ clearly deviates from a $T$-linear behavior, and the $T^2$ model also fails to capture $\Delta\lambda(T)$,  indicating that there are no nodes in the superconducting gap structure of CaSb$_2$. 

The low-temperature $\Delta\lambda(T)$ data are also fit to an exponential temperature dependence (solid red line), which corresponds to a $s$-wave superconducting gap, with
\begin{equation}
\Delta\lambda(T)~\approx~\lambda(0)\sqrt{\frac{\pi\Delta_0}{2T}}\exp(-\frac{\Delta(0)}{k_{\rm B}T}),
\label{equation1}
\end{equation}
where $\Delta(0)$ is the magnitude of the superconducting gap at zero temperature \cite{36Prozorov_2006}. It can be seen that in contrast to the $T^2$ model, the $s$-wave model gives an excellent description of $\Delta\lambda(T)$ at low temperatures, providing strong evidence for nodeless superconductivity in CaSb$_2$. The derived superconducting gap is found to be $\Delta(0)~=~0.9 k_{\rm B}T_{\rm c}$, which is significantly smaller than the weak-coupling limit of $\Delta(0)=1.76 k_{\rm B}T_{\rm c}$, which can arise from the presence of multiple superconducting gaps. This conclusion is corroborated by the excellent agreement between the fit gap value from low-temperature $\Delta\lambda(T)$ and the small gap obtained in specific heat measurements.

To further elucidate the superconductivity in CaSb$_2$, the superfluid density is calculated through $\rho_{\rm s}(T)$~=~$[\lambda(0)/\lambda(T)]^2$ (Fig.~\ref{Figure4}), using $\lambda(0)=89.5$~nm from Ref.~\cite{56PhysRevB.105.184504}. The obtained $\rho_{\rm s}$ is fit to several different models of the superconducting gap function $\Delta_k$, which determines $\rho_{\rm s}$ through
\begin{equation}
\rho_{\rm s}(T) = 1 + 2 \left\langle\int_{\Delta}^{\infty}\frac{E{\rm d}E}{\sqrt{E^2-\Delta_k^2}}\frac{\partial f}{\partial E}\right\rangle_{\rm FS},
\label{equation2}
\end{equation}
where $f(E,T)=[1+\exp({E/k_{\rm B}T)}]^{-1}$ is the Fermi-Dirac distribution and $<...>_{\rm FS}$ represents an average over the Fermi surface \cite{36Prozorov_2006}. The gap function is defined as $\Delta_k(T)=\Delta(T)g_k$, where $g_k$ represents the angular dependence of the gap structure, with $g_k$=1, $\sin \theta$ and $\cos 2\phi$,  corresponding to $s$-, $p$-, and $d$-wave superconducting gaps ($\theta$ is the polar angle and $\phi$ is the azimuthal angle), respectively. The temperature dependence of $\Delta (T)$ is approximated as \cite{37CARRINGTON2003205}:
\begin{equation}
\Delta(T)~=~\Delta(0){\rm tanh}\left\{1.82\left[1.018\left(T_{\rm c}/T-1\right)\right]^{0.51}\right\},
\label{equation3}
\end{equation}  
where $\Delta (0)$ represents the gap magnitude at zero temperature. 

As shown in Fig.~\ref{Figure4}, the experimental normalized superfluid density is not fully captured by any of the $s$-, $p$-, and $d$-wave models. On the other hand, we find that a two-gap $s$-wave model describes the data well over the entire temperature range [solid line in Fig.\ref{Figure4}(a)], where the superfluid density is given by $\rho_{\rm s}~=~x\rho_1(T)+(1-x)\rho_2(T)$, with $\rho_i(T)$ ($i=1,2$) being the superfluid density corresponding to the gap $\Delta_i$ and $x$ is the relative weight of $\Delta_i$. The fit parameters are found to be $\Delta_1(0)=0.84 k_{\rm B}T_{\rm c}$, $\Delta_2(0)=1.83k_{\rm B}T_{\rm c}$, and $x=14\%$, where the small gap $\Delta_1=0.84 k_{\rm B}T_{\rm c}$ is consistent with $\Delta=0.9 k_{\rm B}T_{\rm c}$ derived from the analysis of $\Delta\lambda(T)$ at low temperatures (Fig.~\ref{Figure3}) and the electronic specific heat (Fig.~\ref{Figure2}). 


A multi-gap superconducting state in CaSb$_2$ is consistent with the presence of multiple Fermi surfaces found in both electronic structure calculations and quantum oscillation measurements \cite{18,56PhysRevB.105.184504,31,39PhysRevResearch.3.023086}. While the electronic bands near the Fermi level are dominated by contributions from Sb states, the electron bands crossing the Fermi level along $\Gamma-Z$ and the hole band centered around the $\Gamma$ point are associated with Sb states on distinct sites. Therefore, the presence of multiple Fermi surfaces originating from different orbital states provides a natural explanation for the phenomenology of multi-gap nodeless superconductivity revealed in our measurements. 


As discussed in Ref.~\cite{39PhysRevResearch.3.023086}, since the space group of CaSb$_2$ is $P2_1/m$ (space group No.~11), the associated symmetries include the inversion symmetry $I$, the two-fold screw axis $S_{2y}$, and the mirror symmetry $M_y$. These symmetries give rise to four one-dimensional representations for the superconducting order parameter: $B_g$ ($\chi_I = +1$, $\chi_{M_y} = -1$), $A_g$ ($\chi_I = +1$, $\chi_{M_y} = +1$), $B_u$ ($\chi_I = -1$, $\chi_{M_y} = +1$) and $A_u$ ($\chi_I = -1$, $\chi_{M_y} = -1$), where $\chi_X$ represents the sign of the superconducting order parameter under the symmetry operation $X$. When weak coupling pairing is assumed, the superconducting phases corresponding to different representations can be deduced through the method in Ref.~\cite{39PhysRevResearch.3.023086}.
In particular, for a superconducting order parameter belonging to the $B_g$ representation, it is predicted that CaSb$_2$ has a topologically protected nodal-line superconducting gap structure on the $k_y~=~0$ plane. 
On the other hand, for order parameters belonging to the $A_u$, $B_u$ and $A_g$ representations, there are no symmetry-enforced nodal structures. Superconductivity is topologically nontrivial for $A_u$ and $B_u$ representations, but topologically trivial for the $A_g$ representation. From our experimental results, the superconducting gap structure of CaSb$_2$ is demonstated to be nodeless, which excludes the $B_g$ pairing symmetry.


In addition, for a topological nodal-line semimetal such as CaSb$_2$, there may be degenerate two-dimensional energy bands that are localized on surfaces parallel to the nodal loop, which are called ``drumhead bands"
\cite{48PhysRevB.95.075138,52heikkila2011dimensional,53PhysRevB.84.235126,54PhysRevB.93.205132}. 
It is then possible to generate topological superconductivity from these surface states through the proximity effect via superconductivity in the bulk \cite{1RevModPhys.83.1057,2Sato_2017,6RevModPhys.82.3045,58PhysRevLett.100.096407}. In this scenario, helical spin-polarized electrons on the topological surface state can pair with $p_x + ip_y$ symmetry and realize bound Majorana fermions with non-Abelian statistics. So far, reports on superconductors with nodal lines in the normal state are limited \cite{59,61,62}, and CaSb$_2$ offers a rare opportunity to search for topological superconductivity in a nodal-line semimetal. Further experiments that look for topological surface states and Majorana zero modes in CaSb$_2$ are needed.



\section{Conclusion}
In conclusion, we investigated the superconducting order parameter of the topological nodal-line semimetal CaSb$_2$ through magnetic penetration depth measurements down to 0.07~K. Clear exponential behaviors below $T<T_{\rm c}$ evidence nodeless superconductivity in CaSb$_2$, and further analysis of the superfluid density and electronic specific heat suggest the superconducting gap structure of CaSb$_2$ can be described by a two-gap $s$-wave model, consistent with the presence of multiple Fermi surfaces. These results demonstrate the absence of nodes in the superconducting order parameter of CaSb$_2$ and exclude the $B_g$ representation. Further work is needed to clarify whether CaSb$_2$ realizes topological superconductivity on the surface.

\section{ACKNOWLEDGMENTS}
This work was supported by the National Natural Science Foundation of China (Grant Nos. 11974306, and 12034017), Key R\&D Program of Zhejiang Province, China (Grant No. 2021C01002),  and National Key R\&D Program of China (Grant Nos. 2017YFA0303100).

\bibliography{citationlist}
\end{document}